\documentclass[12pt]{article}
\usepackage{amsmath}
\usepackage{amssymb}
\usepackage{amsfonts}
\usepackage{array}
\usepackage{color}
\usepackage{ulem}
\usepackage{graphicx}
\usepackage{hyperref}
\usepackage{cite}

\begin{document}
\vspace{4cm}
\hspace{7cm}
\begin{center}
{\LARGE\bf Aligned Natural Inflation:\\[2mm]
Monodromies of two Axions}\\[12mm]
{\large Rolf Kappl, Sven Krippendorf and Hans Peter Nilles}\\[6mm]
{\it Bethe Center for Theoretical Physics\\
and\\
Physikalisches Institut der Universit\"at Bonn\\
Nussallee 12, 53115 Bonn, Germany\\[5mm]
}
{\tt kappl, krippendorf, nilles at th.physik.uni-bonn.de}

\begin{abstract}

Natural (axionic) inflation~\cite{Freese:1990rb} can 
 accommodate sizeable primordial tensor modes but suffers
from the necessity of trans-Planckian variations of the inflaton field. This problem can be 
solved via the mechanism of aligned axions ~\cite{Kim:2004rp}, 
where the aligned axion spirals down in the
potential of other axions. We elaborate on the mechanism in view of the recently
reported observations of the BICEP2 collaboration~\cite{{Ade:2014xna}}.

\end{abstract}
\vspace{2cm}
\includegraphics[width=0.5\textwidth]{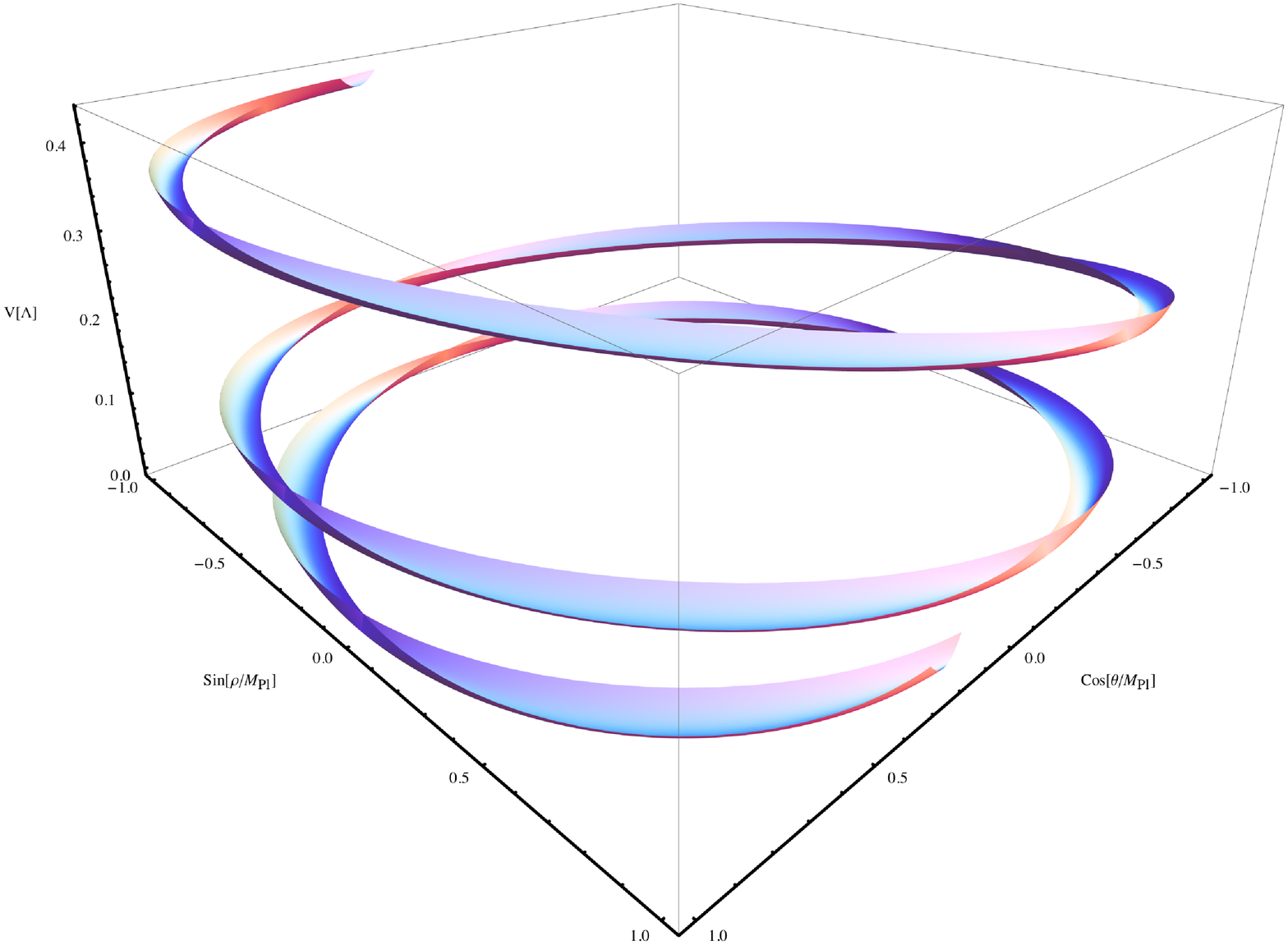}\\[0.7cm]
{\bf The motion of the inflaton:}\\
{\footnotesize The aligned axion roles down in the valley of a second axion.}
\end{center}
\newpage
\section{Introduction}
Natural inflation~\cite{Freese:1990rb,Adams:1992bn} is one of the best motivated scenarios to describe the
inflationary
expansion  in the early universe. The flatness of the potential is protected by a
(shift)
symmetry modelled after the QCD axion~\cite{Peccei:1977hh}, although at different mass scales. The axion
exhibits a shift symmetry that is perturbatively exact, but potentially broken by non-perturbative (instantonic) effects. In its simplest form~\cite{Freese:1990rb} natural (axionic)
inflation is
derived from the potential 
\begin{equation}
\label{equa:natural}
V (\phi) = \Lambda^{4} \left[1 - \cos \left(\frac{\phi}{f}\right)\right],
\end{equation}
where $f$ is the axion decay constant and $\Lambda$ is the overall scale of the potential. This is a periodic potential with period $0
\leq \phi
\leq 2 \pi f$. If inflation occurs for field values $\phi \ll f$ this coincides with
chaotic
inflation~\cite{Linde:1983gd} in its quadratic form $V (\phi) \approx m^{2} \phi^{2}$. Axions are
well
motivated also in the framework of string theory and its various antisymmetric tensor
fields. Their specific properties as pseudoscalar fields allow a wide spectrum of
potential applications in particle physics and cosmology (see for instance~\cite{Chatzistavrakidis:2012bb}).

With the recent observation of the BICEP2 experiment~\cite{Ade:2014xna}, axionic inflation returned
to
the focus of the discussion~\cite{Freese:2014nla}, 
as it allows sizeable tensor modes up to a value of $
r
\approx 0.18$. The results of BICEP2 point to a value of $r  = 0.2^{+0.07}_{-0.05}$ (after dust reduction to $r=0.16^{+0.06}_{-0.05}$),  a
much higher value than expected compared to the results obtained by the
Planck satellite~\cite{Ade:2013uln}. More observations are needed to determine the precise value for
$r$. Up to now the results seem to be compatible with the prediction of axionic
inflation.

One general aspect of inflationary models with large tensor modes is the appearance
of trans-Planckian values of the inflaton field known as the Lyth bound~\cite{Lyth:1996im,Lyth:2014yya}. For example, the model
of
quadratic inflation $(V (\phi) = \frac{1}{2} m^{2} \phi^{2})$  with sufficient number of
e-folds
requires the displacement $\triangle \phi \approx 15 M_{\text{Pl}}$, where $M_{\text{Pl}} = 2.4 \cdot 10^{18}
\ \text{GeV}$. We thus have to worry whether our low energy effective description based on
classical gravity is still valid under these circumstances. Quantum gravitational
effects might destroy the properties of the inflationary potential.

This is a generic problem of all models with ``large" tensor modes and we have to find
arguments to assure the flatness of the potential even at trans-Planckian values of the
field. In absence of a complete theory of quantum gravity this is a severe problem.
Axionic inflation has an ingredient that helps in this direction: its shift symmetry
might be respected by such effects. This is supported by string theoretic arguments
through the appearance of (discrete) gauge symmetries that survive in the ultra-violet
completion~\cite{Lebedev:2007hv,Kappl:2008ie,Nilles:2008gq,BerasaluceGonzalez:2011wy,Kim:2014dba}.

In that sense, axionic inflation appears as the most attractive
scenario
to avoid the problems of trans-Planckian field values, although these questions have to
be analysed on a case by case basis.\footnote{Discrete symmetries in string constructions have been discussed for example in
refs.~\cite{Kobayashi:2006wq,Choi:2009jt,BerasaluceGonzalez:2011wy,Nilles:2012cy,Ibanez:2012wg,BerasaluceGonzalez:2012vb,BerasaluceGonzalez:2012zn,Anastasopoulos:2012zu,Honecker:2013hda,Marchesano:2013ega,Kim:2013bla,Berasaluce-Gonzalez:2013bba,Nilles:2014owa}.}

So let us concentrate on axionic inflation. Unfortunately we still have to face some
problems. In the
potential (\ref{equa:natural}), trans-Planckian values of the inflation field $\triangle \phi \geq
M_{\text{Pl}}$
require $2 \pi f \geq M_{\text{Pl}}$. In fact, in the region of high tensor modes
$r \geq 0.1$ we obtain $f \geq M_{\text{Pl}}$ and this poses a severe problem.
Within the framework of string theory we expect $f $ to be at most as large as the
string scale $M_{\text{String}} \leq M_{\text{Pl}}$~\cite{Choi:1985je,Banks:2003sx,Svrcek:2006yi}.\footnote{Because of the rather high value of the observed tensor modes we expect large values
for the potential
$ V^{1/4}~\sim~2\cdot~10^{16}~{\rm GeV}$ and a rather high value $M_{\text{String}}\geq  V^{1/4}$
of the string scale as well.}
Values of $M_{\text{String}}$ and $f$ larger than $M_{\text{Pl}}$ would lead us to the uncontrollable
regime of strongly interacting string theory where a meaningful discussion of the
flatness of a potential will be impossible. A simple picture of axionic inflation with
a single axion and potential (\ref{equa:natural}) is therefore problematic.

As we have discussed earlier, string theory might provide various axion candidates, we
would have the option to consider models with various axions, and this is what we want
to explore in this paper. We shall follow the suggestion of ref.~\cite{Kim:2004rp} of
axionic inflation with aligned axions: one axion spiralling down in the valley of a
second one (see Figure~\ref{fig:spiral} and its copy on the title page). This mechanism encodes all the nice features of schemes later called
``axion monodromy''~\cite{Silverstein:2008sg,McAllister:2008hb} in a somewhat different set-up (see also~\cite{Pajer:2013fsa} for an overview on axion inflation).
The alternative suggestion of many non-interacting axions, N-flation~\cite{Dimopoulos:2005ac}
differs significantly from the mechanism described here. In addition it requires a really
large number of axion fields~\cite{Kim:2006ys} that might be problematic in an explicit
realisation. We shall comment on these mechanisms later in this paper (Section~\ref{sec:discussion}).

\section{A two axion field theoretical realisation}
\label{sec:model}

To illustrate axionic inflation with aligned axions, let us recall the realisation presented by Kim, Nilles and Peloso in~\cite{Kim:2004rp}.
 We start with the following two field model
\begin{eqnarray}
{\cal L}(\theta,\rho)&=&(\partial\theta)^2+(\partial\rho)^2-V(\rho,\theta)\ , \label{eq:lag}\\
V(\theta,\rho)&=&\Lambda^4 \left(2-\cos{\left(\frac{\theta}{f_1}+\frac{\rho}{g_1}\right)}-\cos{\left(\frac{\theta}{f_2}+\frac{\rho}{g_2}\right)}\right). \label{eq:pot}
\end{eqnarray}
Here we introduce no different energy scales for the cosine factors, which do not alter the results of our following discussion. There is a flat direction in this potential, if the following relation holds
\begin{equation}
\label{equa:relation}
\frac{f_1}{g_1}=\frac{f_2}{g_2}\, .
\end{equation}
This can be seen by looking at the two mass eigenvalues near the origin ($\theta=\rho=0$)
\begin{equation}
\lambda_{1/2}=F\pm\sqrt{F^2+\frac{2g_1g_2f_1f_2-f_2^2g_1^2-f_1^2g_2^2}{ f_1^2f_2^2f_1^2g_2^2}}\ ,\;F=\frac{g_1^2g_2^2(f_1^2+f_2^2)+f_1^2f_2^2(g_1^2+g_2^2)}{2f_1^2f_2^2g_1^2g_2^2}
\label{equa:eigenvalues}
\end{equation}
where the last term under the square root simplifies to zero if relation (\ref{equa:relation}) holds and one eigenvalue remains zero.

\begin{figure}
\includegraphics[width=0.33\textwidth]{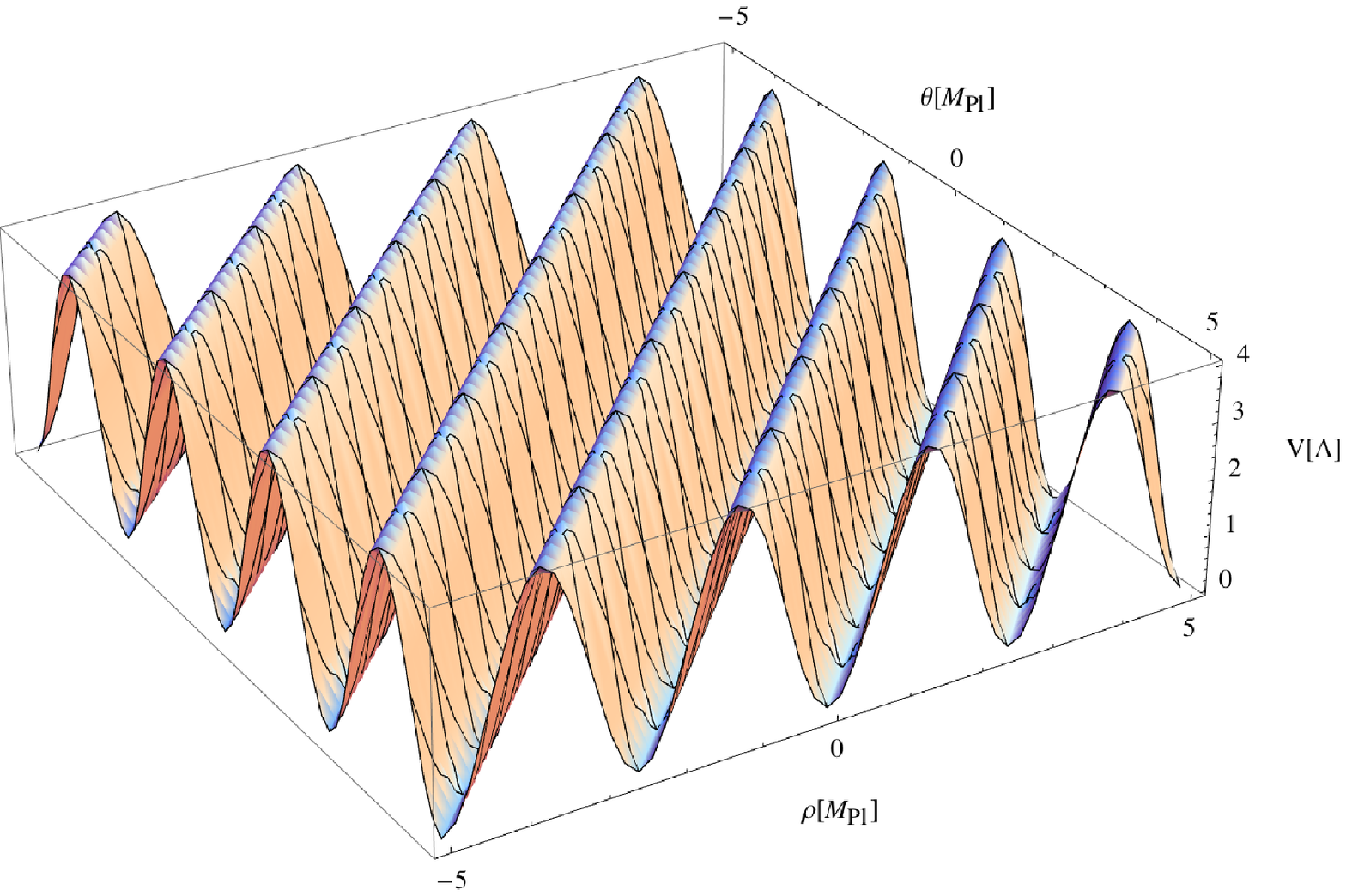}
\includegraphics[width=0.33\textwidth]{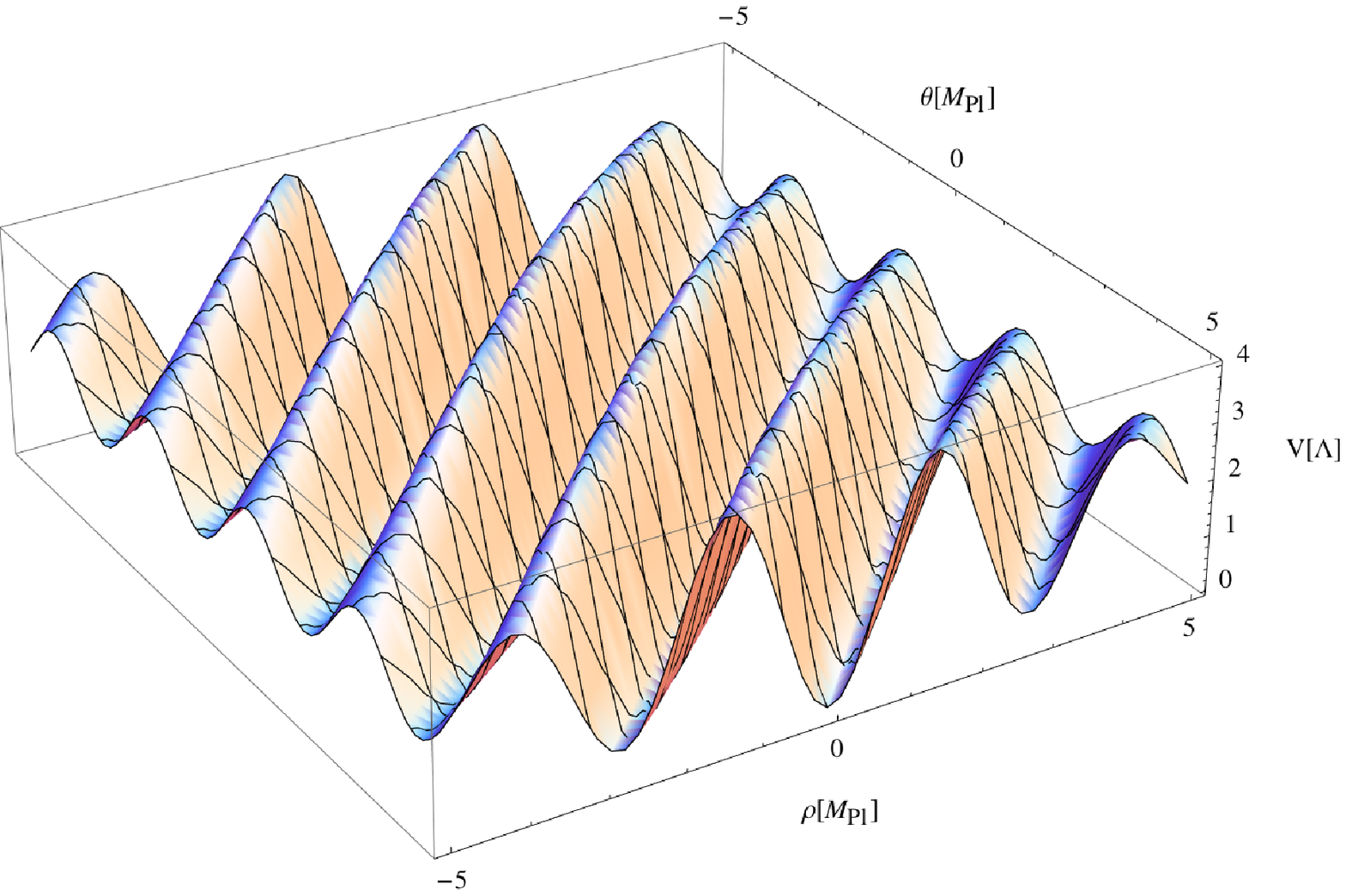}
\includegraphics[width=0.33\textwidth]{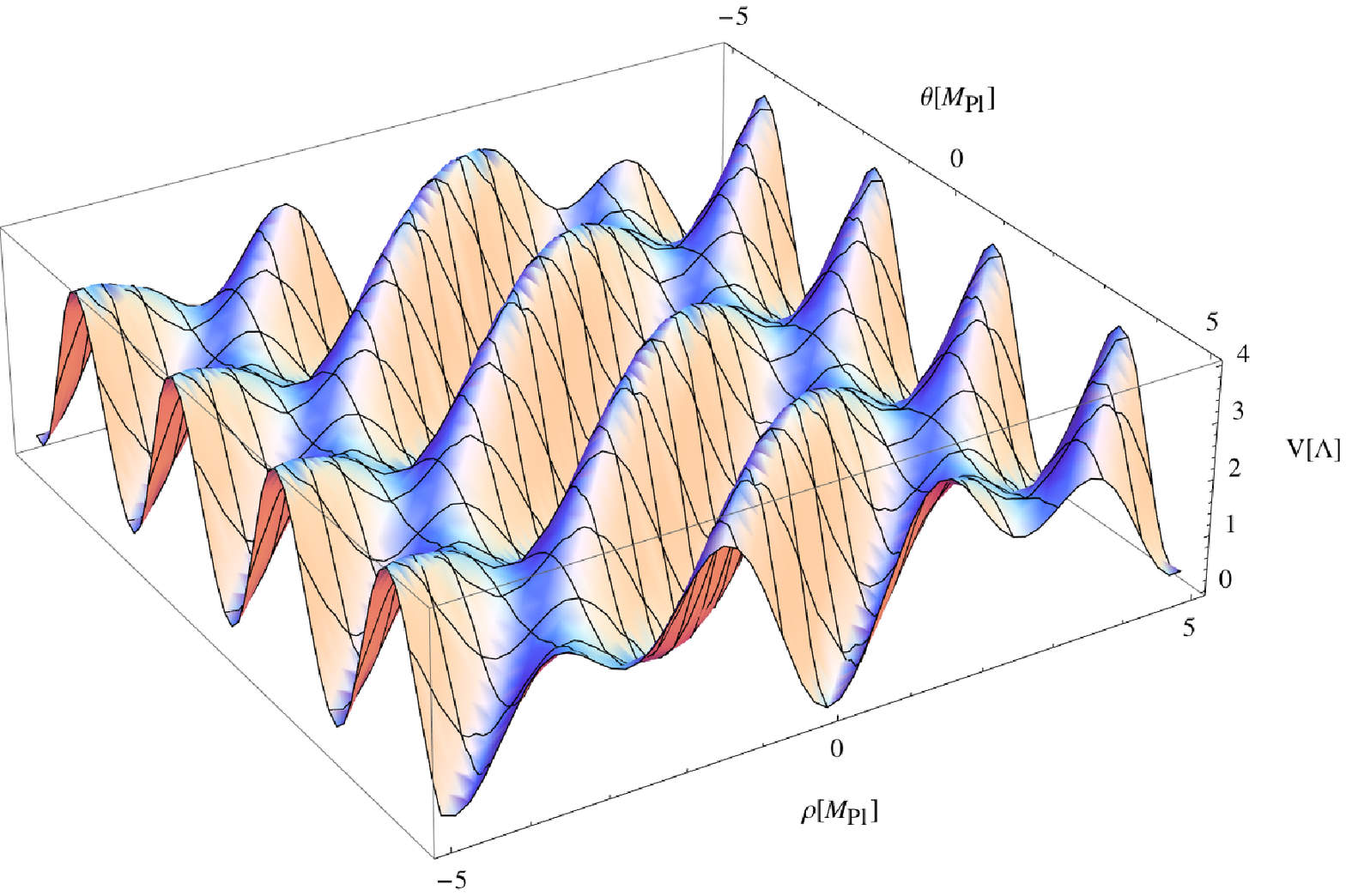}
\caption{\footnotesize{The potential for parameters $f_1=f_2=g_1=0.5 M_{\text{Pl}}$ and the values 
$\alpha=0,$ $\alpha=0.1,$ and $\alpha=0.3.$} \label{fig:potplots}}
\end{figure}

We are interested in breaking this flat direction by the small parameter $\alpha,$ which we define as follows:
\begin{equation}
g_2=\frac{f_2}{f_1}g_1+\alpha\, .
\label{eq:expansion}
\end{equation}
We would like to utilise this flat direction for inflation, while keeping the other direction fixed. More precisely, we are interested in obtaining inflation near the origin. A visualisation of the potential with a universal decay constant $\hat{f}=f_1=f_2=g_1$ and the flat direction can be found in Figure~\ref{fig:potplots} which shows the potential for various values of $\alpha.$  

The first step to analyse inflation, is to go into a suitable basis which parametrises our inflationary trajectory. To set notation we introduce the following new coordinates:
\begin{equation}
(\theta,\rho)\to (\xi, \psi)\ ,
\end{equation}
where $\xi$ parametrises the flattest direction in the potential around the origin and $\psi$ its orthogonal direction. This flattest direction is given by the smallest eigenvalue of the Hessian of the potential $V(\theta,\rho)$ in~\eqref{eq:pot} (see equation (\ref{equa:eigenvalues})). Let us denote the eigenvectors of the Hessian by $v_1$ and $v_2$ with $v_1$ corresponding to the smaller eigenvalue $\lambda_1.$ The transformation is then explicitly given by
\begin{equation}
\left(\begin{array}{c}
\xi\\
\psi
\end{array}\right)=
\frac{1}{\det{\hat{M}}}\left(\begin{array}{cc}
|v_2|\ v_{11}&|v_2|\ v_{12}\\
|v_1|\ v_{21}&|v_1|\ v_{22}
\end{array}
\right)
\left(\begin{array}{c}
\theta\\
\rho
\end{array}\right)\equiv M \left(\begin{array}{c}
\theta\\
\rho
\end{array}\right),
\end{equation}
where $v_i=(v_{i1},v_{i2})^T$ with \(i\in\{1,2\}\) and $\hat{M}$ denotes the matrix out of eigenvectors. The normalisation of this transformation is chosen that the new basis $(\xi,\psi)$ has canonical kinetic terms. The inverse transformation is given by
\begin{equation}
M^{-1}=\left(\begin{array}{c c}
\frac{v_{22}}{|v_2|} & -\frac{v_{12}}{|v_1|}\\
-\frac{v_{21}}{|v_2|} & \frac{v_{11}}{|v_1|}
\end{array}\right)\textcolor{blue}{.}
\end{equation}

Note that this field transformation as of now has been model-independent and it can be applied to other models with similar flat directions. Now, turning to the field theoretical model in~\eqref{eq:lag} with the approximation as in~\eqref{eq:expansion}, we find the following expressions for the transformation up to first order in $\alpha$
\begin{eqnarray}
M&=&\frac{1}{\sqrt{f_1^2+g_1^2}}\left(\begin{array}{c c}
f_1 & -g_1\\
-g_1 & -f_1
\end{array}\right)+\frac{\alpha f_1^4}{(f_1^2 f_2+f_2^3)(f_1^2+g_1^2)^{3/2}}\left(\begin{array}{c c}
-g_1 & -f_1\\
-f_1 & g_1
\end{array}\right),\\
M^{-1}&=&\frac{1}{\sqrt{f_1^2+g_1^2}}\left(\begin{array}{c c}
f_1 & -g_1\\
-g_1 & -f_1
\end{array}\right)+\frac{\alpha f_1^4}{(f_1^2 f_2+f_2^3)(f_1^2+g_1^2)^{3/2}}\left(\begin{array}{c c}
-g_1 & -f_1\\
-f_1 & g_1
\end{array}\right).
\end{eqnarray}

\begin{figure}
\begin{center}
\includegraphics[width=0.5\textwidth]{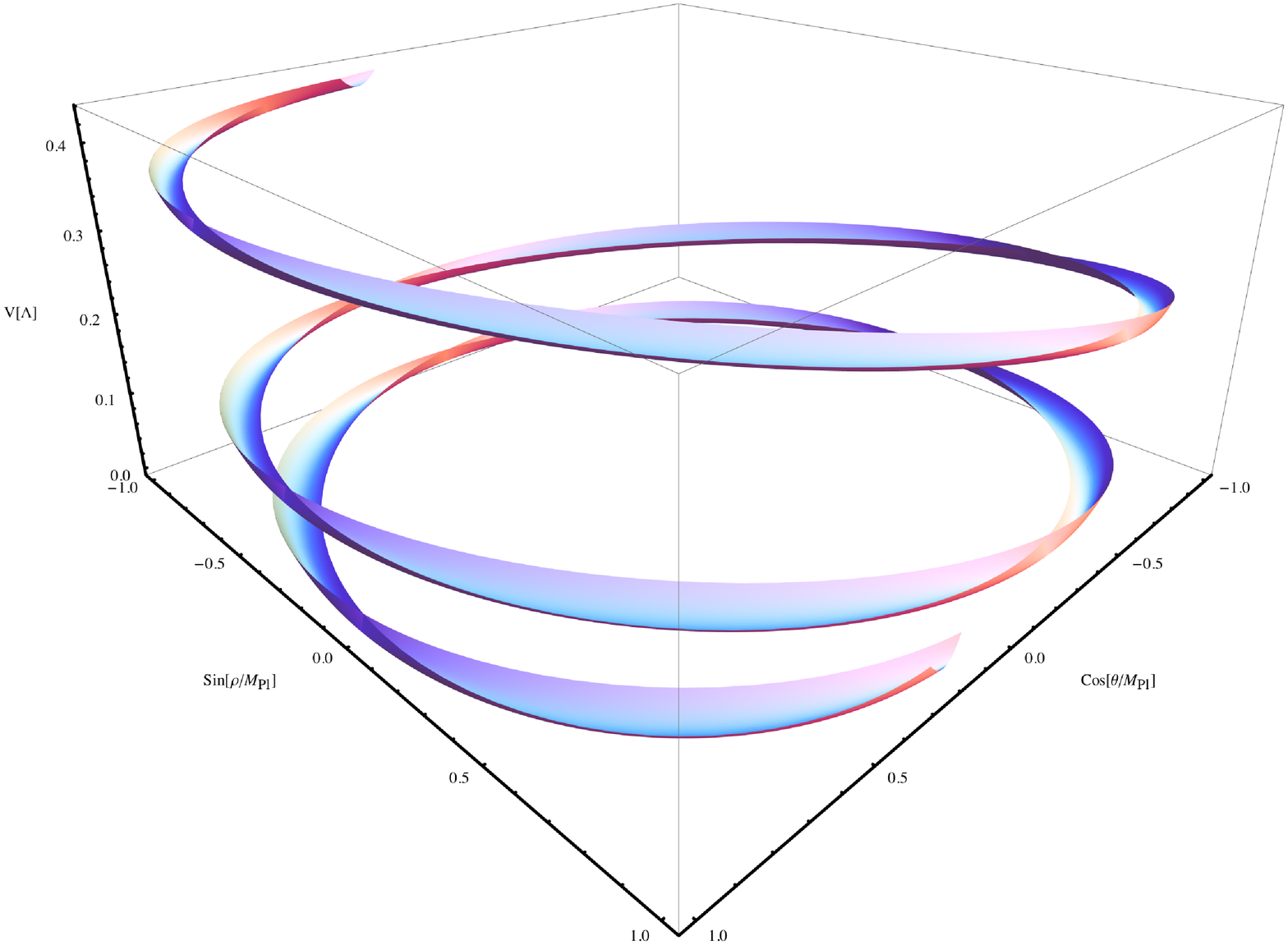}
\end{center}
\caption{\footnotesize{Parametric plot of the potential for parameters $f_1=f_2=g_1=0.5 M_{\text{Pl}}$ and $\alpha=0.1$ starting from $\xi=10.$ The $(x,y)$-coordinates are parametrised by $\cos{(\theta/M_{\rm Pl})}$ and $\sin{(\rho/M_{\rm Pl})}$ to visualise the axionic shift symmetries in the $\theta$ and $\rho$ directions.}\label{fig:spiral}}
\end{figure}

We are interested in parameter ranges in the above potential~\eqref{eq:pot} which are suitable for inflation. Here we focus on generating inflation along the `flat' direction parametrised by $(\xi,\psi=0)$ in the neighbourhood of the origin. 
To analyse the inflationary foot-prints of this particular single field model, let us re-write the potential as follows:
\begin{equation}
V(\xi)=\Lambda^4\left[2-\cos{\left(m_1(f_i,g_1,\alpha) \xi\right)}-\cos{\left(m_2(f_i,g_1,\alpha) \xi\right)}\right].
\end{equation}
The coefficients $m_i$ are to first order in $\alpha$ given by:
\begin{eqnarray}
m_1(f_i,g_1,\alpha)&=&-\frac{\alpha  f_1^3  }{\sqrt{f_1^2+g_1^2} \left(f_1^2 f_2 g_1+f_2^3 g_1\right)}+ \frac{\alpha ^2 f_1^4   \left(f_1^2 \left(2 f_2^2+g_1^2\right)+f_2^2 g_1^2\right)}{f_2^2 g_1^2 \left(f_1^2+f_2^2\right)^2 \left(f_1^2+g_1^2\right)^{3/2}},\\
m_2(f_i,g_1,\alpha)&=&\frac{\alpha  f_1^2  }{\sqrt{f_1^2+g_1^2} \left(f_1^2 g_1+f_2^2 g_1\right)}+\frac{\alpha ^2 f_1^3   \left(f_1^4-f_1^2 \left(f_2^2+g_1^2\right)-f_2^2 g_1^2\right)}{f_2 g_1^2 \left(f_1^2+f_2^2\right)^2 \left(f_1^2+g_1^2\right)^{3/2}}.
\end{eqnarray}
The leading order expansion in $\xi$ matches with the expansion of a potential as in natural inflation
\begin{equation}
V(\xi)=\Lambda^4\left[1-\cos{\left(\frac{\xi}{f_{\text{eff}}}\right)}\right]\, ,
\end{equation}
where the effective decay constant is given by
\begin{equation}
f_{\text{eff}}=\frac{f_2 g_1 \sqrt{(f_1^2+f_2^2) (f_1^2+g_1^2)}}{f_1^2 \alpha}\, .
\end{equation}
We see that by aligning the decay constants appropriately an arbitrarily large effective decay constant can be arranged. A visualisation of the potential along the inflaton direction $\xi$ is shown in Figure~\ref{fig:spiral} where we parametrise the $(x,y)$-coordinates by $\cos{(\theta/M_{\rm Pl})}$ and $\sin{(\rho/M_{\rm Pl})}$ to visualise the axionic shift symmetries in the $\theta$ and $\rho$ directions.

In this model, the slow-roll parameters are then found to be:
\begin{eqnarray}
\nonumber \epsilon&=&\frac{M_{\text{Pl}}^2}{2}\left(\frac{V'(\xi)}{V(\xi)}\right)^2\\
\nonumber &=&\frac{M_{\text{Pl}}^2}{2}\left(\frac{m_1 \sin (m_1 \xi )+m_2 \sin (m_2 \xi )}{-\cos (m_1 \xi )-\cos (m_2 \xi )+2}\right)^2\\
&\approx& \frac{M_{\text{Pl}}^2}{2}\left(\frac{4}{\xi^2}+\frac{2 \alpha^2f_1^4 (f_1^4 - 12 f_1^2 f_2^2 + f_2^4)}{3 f_2^2 (f_1^2 +  f_2^2)^3 g_1^2 (f_1^2 + g_1^2)}+{\mathcal O}(\xi^2)\right),\\
\nonumber \eta&=&M_{\text{Pl}}^2\frac{V''(\xi)}{V(\xi)}\\
\nonumber &=&M_{\text{Pl}}^2\frac{m_1^2 \cos (m_1 \xi )+m_2^2 \cos (m_2 \xi )}{-\cos (m_1 \xi )-\cos (m_2 \xi )+2}\\
&\approx& M_{\text{Pl}}^2\left(\frac{2}{\xi^2}+\frac{5 \alpha^2f_1^4 (f_1^4 - 12 f_1^2 f_2^2 + f_2^4)}{6 f_2^2 (f_1^2 +  f_2^2)^3 g_1^2 (f_1^2 + g_1^2)}+{\mathcal O}(\xi^2)\right),\\
n_s&=&1-6\epsilon+2 \eta\ ,\\ 
r&=&16\epsilon\, .
\end{eqnarray}
In the respective last lines we provide the leading order expressions in the expansions of $\alpha$ and $\xi$ around zero.

\begin{figure}
\begin{center}
\includegraphics[width=0.55\textwidth]{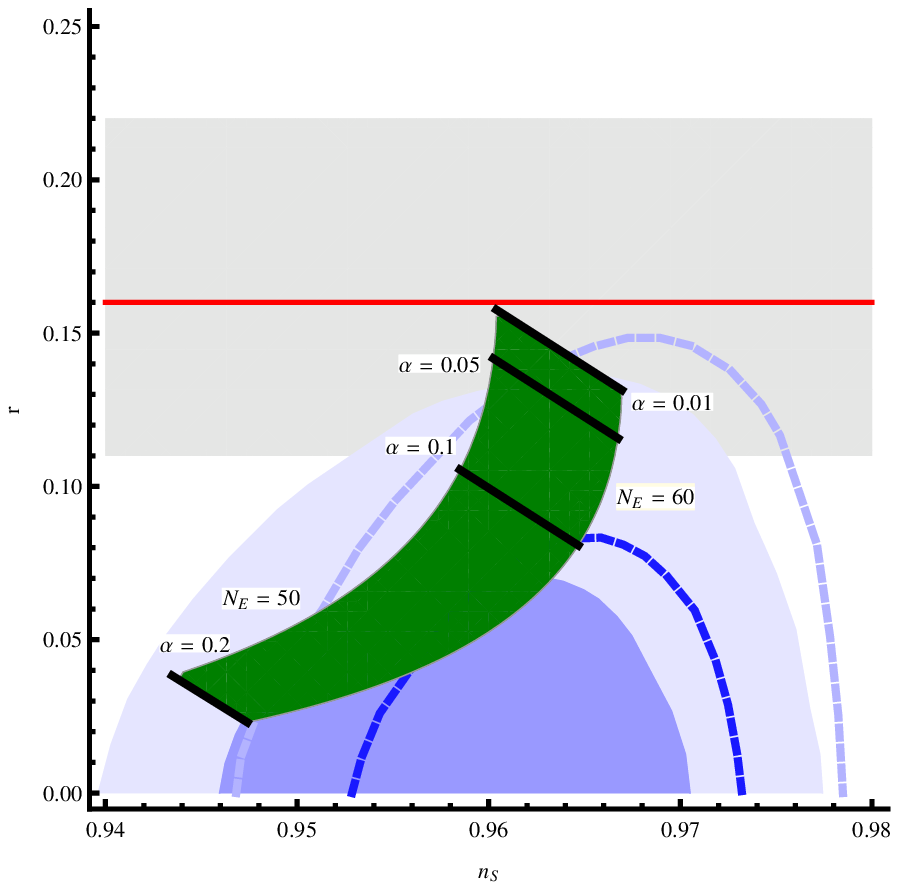}\\
\includegraphics[width=0.35\textwidth]{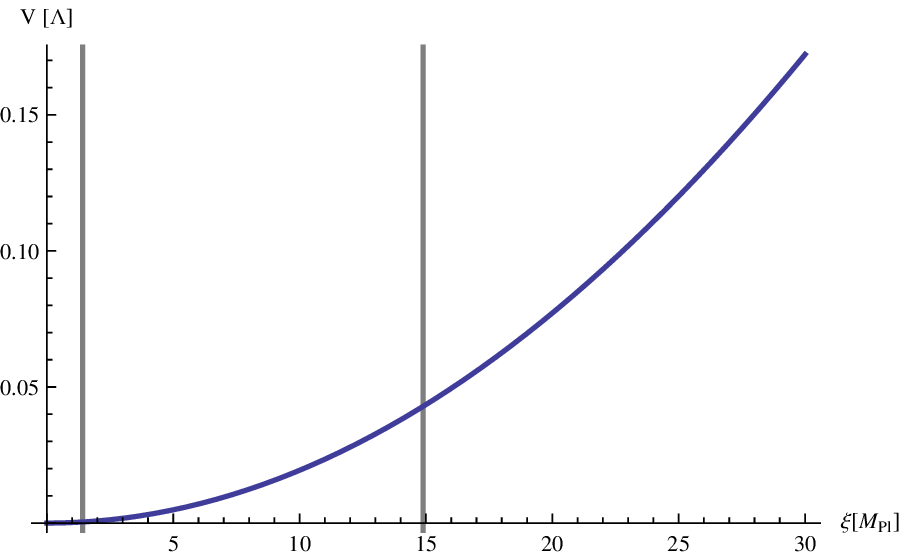}
\includegraphics[width=0.35\textwidth]{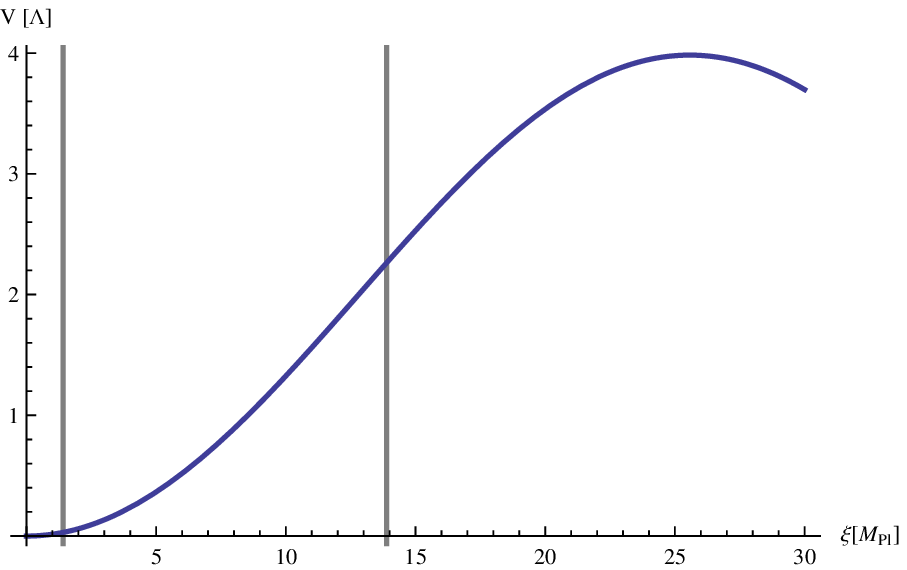}
\end{center}
\caption{ \footnotesize{$n_s-r-$plane profile for aligned natural inflation. The green band shows the parameter-region which is obtained in the aligned natural inflation model for parameters $\alpha=0.01\ldots 0.2 M_{\rm Pl},$ $f_1=f_2=g_1=0.5 M_{\rm Pl}$ and $N_E=50\ldots 60.$ The red line corresponds to the measured value by BICEP2 after reducing dust, the grey band indicates the 
68 percent confidence level~\cite{Ade:2014xna}. The blue regions correspond to the parameter space allowed by combined data from Planck, WMAP polarisation, lensing, ACT and SPT at the 68 and 95 percent confidence level~\cite{Ade:2013uln}. The dotted regions correspond to the parameter space allowed by combined data from Planck, WMAP polarisation, lensing and BAO at the 68 and 95 percent confidence level~\cite{Ade:2013uln}. The potential plots show the range where inflation takes place for $\alpha=0.01$ and $\alpha=0.1.$}\label{fig:nsrplane}}
\end{figure}

Equipped with these slow-roll parameters, we can now determine the model footprint in the $n_s-r-$plane. The result is shown in Figure~\ref{fig:nsrplane} for varying values of the expansion parameter $\alpha.$ Note that by varying the other parameters $f_{1,2}$ and $g_1$ the model footprint remains the same. We find that the footprint is very similar to the one of natural inflation (e.g.~\cite{Freese:2014nla}). As outlined in recent analysis on natural inflation, this is a very attractive area of parameter space in the light of the recent BICEP2 observation of tensor modes. Note also that the same region of natural inflation can be achieved by bottom-up extensions of natural inflation (see for instance the recent papers~\cite{Czerny:2014qqa,Antusch:2014cpa,McDonald:2014oza}).

\section{Discussion}
\label{sec:discussion}

We have seen that the mechanism of aligned axions solves the problem of
``trans-Planckian" field values and thus completes the scenario of natural (axionic)
inflation~\cite{Kim:2004rp}. It is characterised by a new parameter $\alpha$ that measures the
amount of alignment ($\alpha=0$ corresponds to full alignment). A precise measurement of the
tensor mode ratio $r$ would allow us to determine (within this scheme) a narrow range
for the value of $\alpha$ (see Figure~\ref{fig:nsrplane}). Natural inflation is limited to $r\leq 0.18$ and for 
these values of $r,$ $\alpha$ has to be small. Values of $r\sim 0.1$ would correspond
to $\alpha/\hat{f}\approx 0.2$, a rather light alignment. Therefore we are confident that such values
can be achieved in explicit models without a strongly fine-tuned value of $\alpha$. 

Before we enter the details of potential model building let us comment on some alternative
suggestions for the solution of the trans-Planckian problem. One of them is so-called
N-flation~\cite{Dimopoulos:2005ac}, motivated by the mechanism of assisted inflation~\cite{Liddle:1998jc}. It should not be confused with the alignment mechanism
discussed in this paper as it suggests the existence of $N$ {\it non-interactive} axions.
With such a setup one can construct an effective axion scale $f_{\rm eff}\sim \sqrt{N} f_i$
(with individual scales $f_i$). In a realistic setting this mechanism requires many axions
${\mathcal O}(10^3)$ (see for instance~\cite{Kim:2006ys,Cicoli:2014sva,Baumann:2014nda}). The existence of many fields leads
to a renormalisation of the Planck scale proportional to $\sqrt N$ as well and it is not clear
whether $f_{\rm eff}$ could keep up with the ``increase" of the Planck-mass.

Other suggestions for a solution of the trans-Planckian problem use mechanism very
similar to the alignment mechanism of KNP~\cite{Kim:2004rp}. They are based on axionic fields in
specific brane backgrounds motivated from string theory~\cite{Grimm:2007hs,Silverstein:2008sg,McAllister:2008hb,Hebecker:2014eua,Palti:2014kza,Marchesano:2014mla,Higaki:2014pja,Tye:2014tja} and in F-theory~\cite{Blumenhagen:2014gta,Grimm:2014vva}. These brane backgrounds (as e.g. NS5-branes)
break the axionic symmetries explicitly and provide a slope for the axion to slide down.
With this breakdown of the axionic symmetry one might worry whether the original
shift symmetry still protects trans-Planckian displacements. The effect of 
brane-backreactions might be too strong for the mechanism to survive~\cite{Conlon:2011qp,Baumann:2014nda}.
The original shift symmetry might thus not suffice to guarantee the flat potential needed
for inflation. More work needs to be done to see whether these scenarios really solve the
trans-Planckian problem (see~\cite{Germani:2010hd,Harigaya:2014eta} for alternative approaches).
A scheme known as ``Dante's Inferno"~\cite{Berg:2009tg} considers a two-axion scheme as in~\cite{Kim:2004rp} and
adds a background brane to tilt the potential. As this brane might lead to uncontrollable
backreactions as well, it might also be counterproductive to a solution of the
trans-Planckian problem.

Of course, we do not yet have an explicit incorporation of the aligned axion scenario
in string theory, but we think it is worthwhile to invest some work in this direction. In the
original approach~\cite{Kim:2004rp} one was considering the axions from the NS-NS two form $B_2$
in the heterotic string with a breakdown of the shift symmetry through gauge instantons.
There it was rather difficult to obtain a small value of $\alpha$. As is known by now,
the NS-NS axions might also lead to a so-called $\eta$-problem
through moduli mixing in the K\"ahler potential  (see for instance~\cite{Baumann:2014nda}). At this point it seems that the $C$-axions
for instance from the R-R two form $C_2$ might be better suited for our purpose. There are various
effects that break the shift symmetries in a controllable way, that might allow for realistic 
values of the alignment parameter $\alpha$ without specific fine-tuning. In any case, let 
us first wait for a more precise measurement of the value of $r$ that determines the
required value of $\alpha$.

\subsection*{Note added}

During our work on the alignment mechanism we became aware of other work
that might be relevant in this direction. In~\cite{Choi:2014rja}, the authors discuss extensions of the KNP-mechanism with more than two axions, which can help to ameliorate the potential fine-tuning in $\alpha.$ In~\cite{Kallosh:2014vja}, the embedding of natural inflation in supergravity is discussed based on earlier work in the framework of
supergravity shift symmetries (see for instance~\cite{Gaillard:1995az,Kawasaki:2000ws,BlancoPillado:2004ns,BlancoPillado:2006he}).

\subsection*{Acknowledgements}
This work was supported by the SFB-Transregio TR33 ``The Dark Universe" (Deutsche Forschungsgemeinschaft).

\bibliographystyle{utphys}
\bibliography{bibliography}
\end{document}